\documentclass[%
reprint,
superscriptaddress,
 aps,
 prl,
]{revtex4-2}

\usepackage{amsthm,amsmath,amssymb}
\usepackage{graphicx}
\usepackage{subfig}
\usepackage{dcolumn}
\usepackage{bm}
\usepackage{physics}
\usepackage{color}

\begin{document}


\title{Quantum Reliability}

\author{L.X.Cui}
\affiliation{Beijing Computational Science Research Center, Beijing, 100193, China}

\author{Y-M.Du}
\email{ymdu@gscaep.ac.cn}
\affiliation{Graduate School of China Academy of Engineering Physics, Beijing, 100193, China}

\author{C.P.Sun}
\email{suncp@gscaep.ac.cn}
\affiliation{Graduate School of China Academy of Engineering Physics, Beijing, 100193, China}
\affiliation{School of Physics, Peking University, Beijing, 100871, China}
\affiliation{Beijing Computational Science Research Center, Beijing, 100193, China}

\date{\today}

\begin{abstract}

Quantum technology has led to increasingly sophisticated and complex quantum devices. Assessing their reliability (quantum reliability) is an important issue. Although reliability theory for classical devices has been well developed in industry and technology, a suitable metric on quantum reliability and its loss has not been systematically investigated. Since reliability-loss depends on the process, quantum fidelity does not always fully depict it. This study provides a metric of quantum reliability by shifting the focus from state-distinguishing to trajectory-distinguishing. In contrast to the conventional notion of classical reliability, which is evaluated using probabilistic measurements of binary logical variables, quantum reliability is grounded in the quantum probability amplitude or wave function. This research provides a universal framework for reliability theory encompassing both classical and quantum devices. It offers a new perspective on quantum engineering by elucidating how intensely the real quantum process a device undergoes influences its performance.

\end{abstract}

\maketitle

\emph{Introduction}.---
In industry and technology, reliability plays a crucial role in various fields such as aviation, aerospace, electronics, and major projects. It serves as a fundamental concept for quantifying the ability to accomplish tasks. In the realm of reliability engineering, the duration and effectiveness with which devices maintain their capabilities are of primary concern. Detailed explanations can be found in reliability textbooks  \cite{Mathematical_Theory_of_Reliability,gnedenko2014mathematical}.
Recently, the field of reliability engineering has encountered new challenges that demand a deeper understanding of the physical principles governing the degradation of system reliability\cite{Rocchi}.

One prominent challenge arises from the need for high-integration equipment or devices, where the components are reduced to extremely small sizes. Therefore, the quantum effects between these components cannot be disregarded. Additionally, some devices accomplish their task by consuming the quantum resources in them\cite{Quantumresourcetheories}. Examples include quantum-sensing systems\cite{Quantumsensing}, quantum simulators\cite{QuantumSimulators:ArchitecturesandOpportunities}, and quantum memories\cite{QuantummemoryforphotonsDarkstatepolaritons}. To confront these challenges, a novel framework for reliability analysis based on quantum mechanics is required.
The terms quantum systems and devices in this letter refer to functional quantum systems that rely on quantum coherence and phase.
Firstly, the classical definition of reliability does not always hold for devices with quantum coherence. For instance, a simple device with a single qubit on the Bloch sphere cannot be categorized into reliable and unreliable states due to the inability to distinguish non-orthogonal states\cite{QuantumComputationandQuantumInformation}.
Secondly, a trajectory-depend quantity is needed to describe the reliability-loss(degradation) of quantum systems. Here, the term trajectory refers to the system's state sequence of time, which is explicitly defined later. Traditional metrics like quantum fidelity, which quantifies the discrepancy between quantum states, are often inadequate for evaluating reliability, let alone the reliability loss of systems. For example,  a finer description of the process is more suitable for the system incorporating feedback controls\cite{Wiseman2000,Wiseman2001,Sanders2010}  (for more information, see Supplemental material VII\cite{Supplementarymaterial}). Furthermore, the trajectories for a quantum system can interfere with each other and a measure to differentiate a given trajectory from the target trajectory is rarely considered before. 
Thirdly, achieving higher reliability often necessitates the implementation of fault-tolerant systems. Such systems involve multiple subsystems that are integrated to construct fault-tolerant structures \cite{ProbabilisticLogies, Reliablecircuitsusinglessreliablerelays,ReliabilityofaPhysicalSystem}, including error correction codes for both classical and quantum channels, memories, gates\cite{TheTheoryofErrorcorrectingCodes,Schemeforreducingdecoherenceinquantumcomputermemory,Errorcorrectingcodesinquantumtheory,StabilizerCodesandQuantumErrorCorrection,Quantumerrorcorrectionforquantummemories}, etc.. As we will show below,   the relation between the quantum reliability of a system and that of its components depends on the quantum correlation. 

In this letter, we develop a reliability framework for quantum devices based on the quantum logic\cite{ConsistentQuantumTheory,Consistentinterpretationsofquantummechanics}. 
Since accomplishing tasks for quantum devices is always based on precise operations, 
our goal is to define a measure called quantum reliability to measure how fast a quantum system loses control. The property is given as follows:
quantum reliability is an extension of quantum fidelity\cite{FidelityforMixedQuantumStates,Quantumcoding} that distinguishes a given trajectory from an arbitrary process.   
Quantum reliability also reflects the hierarchy of a quantum device called \emph{structure function}.
The structure function of quantum devices with multiple subsystems is shown to be modified by quantum coherence. A quantum storage system with a fault-tolerant structure is present as an example to illustrate the impact of quantum coherence on system reliability.

\emph{Structure functions of quantum devices}.---
A system is typically composed of various subsystems. 
The connection between these subsystems and the system is depicted by Boolean functions, known as structure functions. However, for quantum systems, the presence of quantum coherence requires more consideration of how to describe the system's structure. 
In this context, we use the terminology of Boolean algebra to characterize the structure of quantum devices. This approach also encompasses the corresponding classical description.


The reliability states (survival and failure) of systems can be represented using the indicator functions on the physical space and the projection operators on the Hilbert space, respectively (see Supplemental materials I\cite{Supplementarymaterial}). The \emph{sample space} $\mathcal{S}$ is composed with all components' reliability states.
Subsets in the sample space are called \emph{events}. All the subsets of $\mathcal{S}$ form a \emph{Boolean algebra} $\mathcal{B}$ under the usual set-theoretic relationships. Events that consist of a single sample are called \emph{elementary} events and the rest are called \emph{compound} events.

For a system with $n$ components, the Boolean algebra of the system is $\mathcal{B}$, and of the components are $\mathcal{B}_1,\cdots, \mathcal{B}_n$. The structure function $\mathfrak{g}$ acts on the events of the components and gets the event in the system of Boolean algebra $\mathcal{B}_1,\cdots, \mathcal{B}_n$ to the Boolean algebra of the system is $\mathcal{B}$, i.e. $\mathfrak{g}: \mathcal{B}_1 \times \mathcal{B}_2 \times \cdots \times \mathcal{B}_n \mapsto \mathcal{B}$. . Events correspond to indicator functions in classical and to projection operators in a quantum fashion.

Suppose the projector of the survival of $i$-th component is $E_i$, and $E_i^{\perp}=I-E_i$ corresponds to the events of failure, where $I$ denotes the identity. The projector $E_{\rm S}$ corresponding to the survival of the system is defined by the structure function $\mathfrak{g}$, which is formally noted as:
\begin{equation}
  E_{\rm S}=\mathfrak{g}(E_1, E_2,\cdots,E_n),
\end{equation}
for example, a logical parallel system: $E_{\rm para}=E_1 \otimes E_2^\perp + E_1^\perp \otimes E_2 +E_1 \otimes E_2$ and a logical series system: $E_{\rm seri}=E_1 \otimes E_2$. The definition of these structures also exists in classical reliability theory  \cite{Mathematical_Theory_of_Reliability}.

There exist some quantum systems where the global Boolean algebra cannot be decomposed into local ones, which is not possible in classical systems. For example, a system consisting of two qubits whose survival space is defined by $E_{\mathrm{S}}=(\ket{00}+\ket{11})(\bra{00}+\bra{11})/2$. One could not divide this system into two one-qubit subsystems according to the reliability logic, despite they are physically segregated.



\emph{Lifetime with quantum measurements}.---
In the context of system reliability, a common concern is the length of time that a system can maintain its survival states, which is referred to as its lifetime.

Let $E(i)$ be the projector for the survival of the system at time $t_i$, and $E^\perp(i)$ for the failure.  Let the initial state of the system be $\ket{\psi}$. Then there are $2^f$ possible sequences of states of the system for a given set of times $t_1<t_2<\cdots <t_f$. The trajectory of the system is described by a train of tensor products of the reliability states of the system at each moment. For example, the tensor product train $\ketbra{\psi} \otimes E(1)\otimes E(2) \otimes E^\perp (3) \cdots\otimes E(f)$ represent for the trajectory $\ket{\psi} \to {\rm survival} \to {\rm survival} \to {\rm failure} \cdots \to {\rm survival}$. Note that the tensor products here act between Hilbert spaces at different time-points of the same system. The \emph{survival trajectory} at time $t_k$ is the trajectory that survived all the previous $k$ moments, written as $ \mathcal{R}_k=E(1)\otimes E(2) \cdots \otimes E(k)$.

For a trajectory $\mathcal{Y}=\ketbra{\psi} \otimes E_1 \cdots  \otimes E_f$
of a closed system, one can define the unitary evolution from time $t'$ to $t$ of the form $U(t',t)$. The \emph{weight}\cite{griffiths1984consistent} of the trajectory $\mathcal{Y}$ is defined as 
\begin{equation}
    \begin{aligned}
        W[\mathcal{Y};U] \equiv {\rm Tr}[
        E_f U(t_f,t_{f-1})E_{f-1}\cdots E_1 U(t_1,t_0)\\
        \ketbra{\psi}U(t_0,t_1)E_1^\dagger \cdots E_{f-1}^\dagger U(t_{f-1},t_f) E_f
        ],
    \end{aligned}
\end{equation}
which is called \emph{survival weight} for a survival trajectory $\mathcal{R}_k$.
The \emph{reliability} $R$ of the system at $t_k$ is defined as the weight of the survival trajectory, $R(k)=W[\mathcal{R}_k,U]$.
One can check that reliability decreases with time, $ R(k+1)\leq R(k).$

The weight could be viewed as an extension of fidelity. To see this, we consider a  weight with two-time points $\mathrm{Tr}[E_{1}U(t_{0},t_{1})E_{0}U^{\dagger}(t_{0},t_{t})]$. This is the state fidelity between the targeting final state and the real final state, and this quantity is used to define the process fidelity with some averages over the initial state \cite{PEDERSEN200747}. Process fidelity regards the process as a whole which only relies on the final state with a given initial state, while quantum reliability relies on the state in the trajectory. Thus the process fidelity, as a two-point quantity, is a weight of the coarse-grained trajectory.  An issue comes from the extension of the definition to mixed states. For the state fidelity, this extension has been well established with maximizing the fidelity between all possible purification. For the weight of mixed trajectory,  that is $E_{i}$ is not a projector but a probability mixture of a number of projectors. This is clear when $\mathrm{Tr}(E_{i})=1$ for all $i$, but for a coarse-grained trajectory  $\mathrm{Tr}(E_{i})=2,3,\cdots$, the purification are not clear yet. An attempt is discussed in Supplemental materials III\cite{Supplementarymaterial}. However, we would like to note that this issue requires further investigation.

The weights assigned here could not always be interpreted as the probability of a trajectory's occurrence. The weights of each trajectory can be interpreted as probabilities only if the trajectories of interest satisfy a consistency condition. 
This is a nature of quantum systems, arising from quantum interference. In classical systems, there is no interference between different trajectories, and thus the consistency condition always holds.

The consistency condition for the survival trajectories is given as follows. 
Consider the trajectory that firstly fails at time $t_{k}$, which is called the failure trajectory at $t_{k}$. Assume that the initial state of the system is $\ketbra{\psi}$. All the trajectories we focus on form a family $\Xi$, including the failure trajectory at time $t_k,\ 1\leq k \leq f$: $\mathcal{F}_k=\ketbra{\psi}\otimes E(1) \cdots \otimes E(k-1) \otimes E^\perp(k)$, and the survival trajectory at time $t_f$: $\mathcal{R}_f=\ketbra{\psi}\otimes E(1)\cdots \otimes E(f)$. The consistency condition \cite{griffiths1984consistent}: for $1\leq k' \leq k \leq f$, the following equation holds,
\begin{equation}
    \begin{aligned}
        \mathrm{Re} \  \mathrm{Tr} [E_k^\perp U(t_k,t_{k-1}) E_{k-1} \cdots E_1 U(t_1,t_0)\ketbra{\psi}\\
        U(t_0,t_1) E_1 \cdots U(t_{k'-1},t_{k'}) E_{k'}^\perp U(t_{k'},t_k)
        ]=0
    \end{aligned}
\end{equation}
For the family satisfies the consistency condition, the weight of each trajectory in the family becomes the probability, i.e., $W[\mathcal{Y};U]\rightarrow {\rm Pr} (\mathcal{Y})$. 

The consistency condition is not held in common. However, one could introduce a measurement apparatus that couples with the system, so that the consistency condition holds.  A measurement protocol is presented as follows.



The Hilbert space of the system and the apparatus is $\mathcal{H}\otimes \mathcal{G}$. Assume that the bases of the apparatus space $\mathcal{G}$ is $\ket{0},\ket{1},\ket{2},\cdots$.
The initial state of the apparatus is $\ketbra{0}$. When the system survives, the apparatus state rises by one, and when the system fails, the apparatus state remains unchanged. This can be done by the measurement operation 
$O_k=E\otimes F_k+ E^\perp \otimes I$
at time $t_k$, where the unitary operators $F_{k}$ could be chosen as $F_k=\ketbra{k}{k-1}+\ketbra{k-1}{k}+\sum_{i\neq k-1,k}\ketbra{i}$.
The survival trajectory of the system and apparatus becomes
\begin{equation}
\mathcal{R}'_k=\ketbra{\psi_0,0}\otimes[E(1)\otimes \ketbra{1}]\cdots \otimes[E(k)\otimes \ketbra{k}].
\end{equation}
One can prove that the family $\{\mathcal{F}'_1,\mathcal{F}'_2,\cdots,\mathcal{F}'_{f},\mathcal{R}'_f\}$ satisfies the consistency condition with $\mathcal{F}'_k\equiv \ketbra{\psi_0,0}\otimes[E(1)\otimes \ketbra{1}]\cdots \otimes[E(k-1)\otimes \ketbra{k-1}] \otimes[E(k)\otimes \ketbra{k}]^\perp$ . The weight of each trajectory is the probability of occurrence of that trajectory, i.e. ${\rm Pr}(\mathcal{R}'_k)=W[\mathcal{R}'_k, U \otimes O]$. It is worth noting that the weights are invariant with measurement, i.e., $W[\mathcal{R}_k,U]=W[\mathcal{R}'_k, U \otimes O]$. However, one can refer to the weights as probabilities only when the measurement is implemented.

The final state of the system and the apparatus is
$\ket{\Phi}=\sum_{k=1}^{f}w_k \ket{\psi_0}\otimes \ket{k}$, where
\begin{equation}
  \begin{aligned}
  w_k = \left(\prod_{j=k+2}^{f} U_j \right) E^\perp (k+1)U_{k+1}\left(\prod_{i=1}^k E(i)U_i \right) ,
  \end{aligned}
\end{equation}
where $U_{k}\equiv U(t_{k-1},t_{k})$ is the evolution operator of the system. The density matrix of the measurement apparatus is 
$
  \rho^\mathrm{M}={\rm Tr_{sys}}\ketbra{\Phi}\equiv \sum_{k,k'\geq 1}\rho^\mathrm{M}_{kk'}\ketbra{k}{k'},$
where $\rho^\mathrm{M}_{kk'}=\mel{\psi_0}{w_{k'}^\dagger w_k}{\psi_0}$.

It is proved that the diagonal elements of the apparatus density matrix are the weights of the failure trajectories $\rho^\mathrm{M}_{kk}=W[\mathcal{F}_k,U]$, and the sum of which gives the survival weight $W[\mathcal{R}_k,U]=1-\sum_{k'=1}^{k}\rho^\mathrm{M}_{k'k'}$.
Define the \emph{lifetime operator} on the apparatus Hilbert space as $\hat{T}=\sum_{k\geq1} k|k\rangle\langle k|$.
Measuring this observable $\mathrm{Tr}[\hat{T}\rho^M]$ leads to the average lifetime of the device.
The apparatus density matrix could be regarded as a quantum version of the lifetime probability density distribution. A classical lifetime probability density distribution is always a function of the failure rate, that is the probability density divided by its cumulative distribution.

After a perfect measurement of trajectory, the apparatus density matrix also contains off-diagonal elements. 
The presence of these off-diagonal elements leads to a different form of entropy in statistical inference, which in turn results in the correction of the inference outcome. Moreover, our case study shows that the apparatus density matrix is a function of the failure rate $x(k)\equiv\rho^{\mathrm{M}}_{kk}/R(k)$, which depends on the coupling Hamiltonian between the environment and the system.

\emph{Example}.---
Next, we consider quantum storage as an illustration.
The quantum storage is made of two-level atoms.
In order to demonstrate the difference between quantum and classical reliability, fault tolerance is implemented with the three-bit-flip code.
The three-bit flip code encodes the one-bit state $\alpha \ket{1}+ \sqrt{1-\alpha^2} \ket{0}$ with the three-bits state $\alpha \ket{111}+ \sqrt{1-\alpha^2} \ket{000}$. 
This encoding is capable of tolerating bit-flip errors of up to one bit and thus be a classical code.
However, the entanglement of such an encoding state implies that the structure function of such a system could be non-classical and depends on the state to be stored. 
In this sense, each physical bit serves as a component. The absence of errors in the physical bit is defined as survival, which corresponds to maintaining the initial state of the input. Meanwhile, the ability to store information correctly, i.e., at most one of the three physical bits has flipped,  corresponds to the survival of the system.
We further assume that the time duration of one measurement could be neglected and the time interval $\delta t$ between any two neighbor measurements is much longer than that of the memory of the environment, thus the Markov approximation holds in such time intervals (see Supplemental materials II\cite{Supplementarymaterial}). During these time intervals, the dynamic of the system is depicted by the Lindblad master equation and three physical bits evolve independently of each other.  


Consider the reliability loss of individual components. The behavior of a two-level atom undergoing spontaneous radiation at a finite temperature can be characterized by the master equation
$
    \dot{\rho} = \gamma_0 (1-N)(\sigma_- \rho \sigma_+ - \frac{1}{2}\{\sigma_+ \sigma_- ,\rho \}) + \gamma_0 N (\sigma_+ \rho \sigma_- - \frac{1}{2}\{ \sigma_- \sigma_+ ,\rho\})
$
where, $\gamma_0$ is the spontaneous emission coefficient  and $0\leq N\leq 0.5$ depends on the temperature of environment $N=1/[\exp( \omega/T)+1]$, $ \ \sigma_-=\sigma_+^{\dagger}=\ketbra{0}{1}$. The initial state of the single qubit is $\ket{\psi_0}$, and the corresponding survival projector is $E=\ketbra{\psi_0}$. The survival trajectory of a single physical qubit is written as $\ketbra{\psi_0}\to E\to E\to \cdots$, and the reliability 
is calculated with Markov approximation,
\begin{equation}
   R_P(t)={\rm Tr}[(P_E \circ\Lambda_{\delta t} ) ^{\circ(t/\delta t)} \ketbra{\psi_0}]\overset{\delta t\rightarrow 0}{=}e^{-M \gamma_0 t},
    \label{eq: r(t)}
\end{equation}
where $(\cdots)^{\circ n}$ denotes $n$ times of function compositions,
$\Lambda_{\delta t}(\rho)=\rho+\dot{\rho} \delta t+O(\delta t^{2}), \ P_E(\rho) = E\rho E$, and $M=\mathrm{Cov}(\sigma_+,\sigma_-)+(1/2-N)\mel{\psi_0}{Z}{\psi_0}$ with $\mathrm{Cov}(A,B)$ being the covariance $\langle\psi_{0}|\{A,B\}|\psi_{0}\rangle/2-\langle\psi_{0}|A|\psi_{0}\rangle\langle\psi_{0}|B|\psi_{0}\rangle$ and $Z$ is the Pauli-Z matrix.

\begin{figure*}[htbp]
\subfloat[]{
    \label{fig: QuantumMemoryWithFaultToleranceStructure} 
    \includegraphics[width=0.3\linewidth]{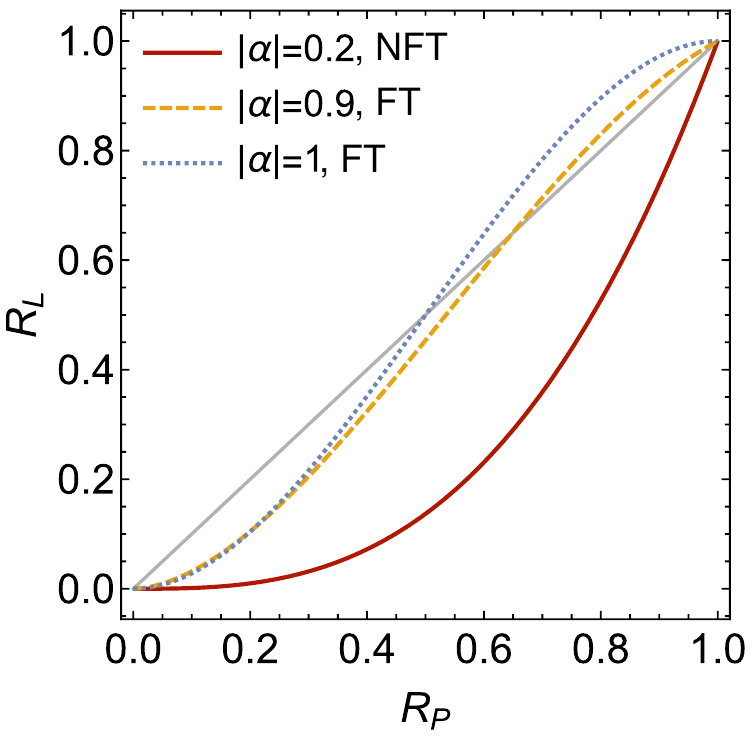}}
\subfloat[]{
    \label{fig: Phase}
    \includegraphics[width=0.3\linewidth]{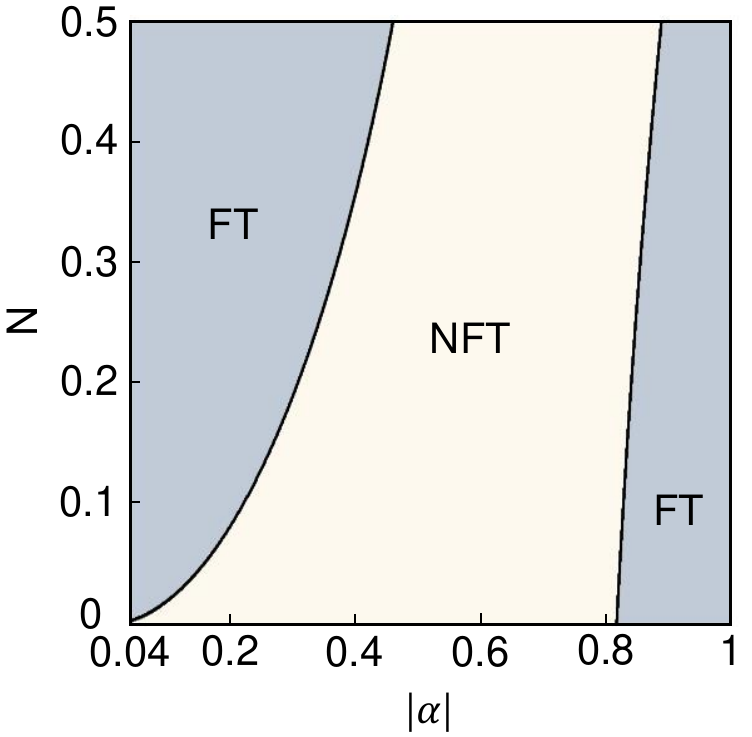}}
\subfloat[]{
    \label{fig: DeltaEntropy}
    \includegraphics[width=0.35\linewidth]{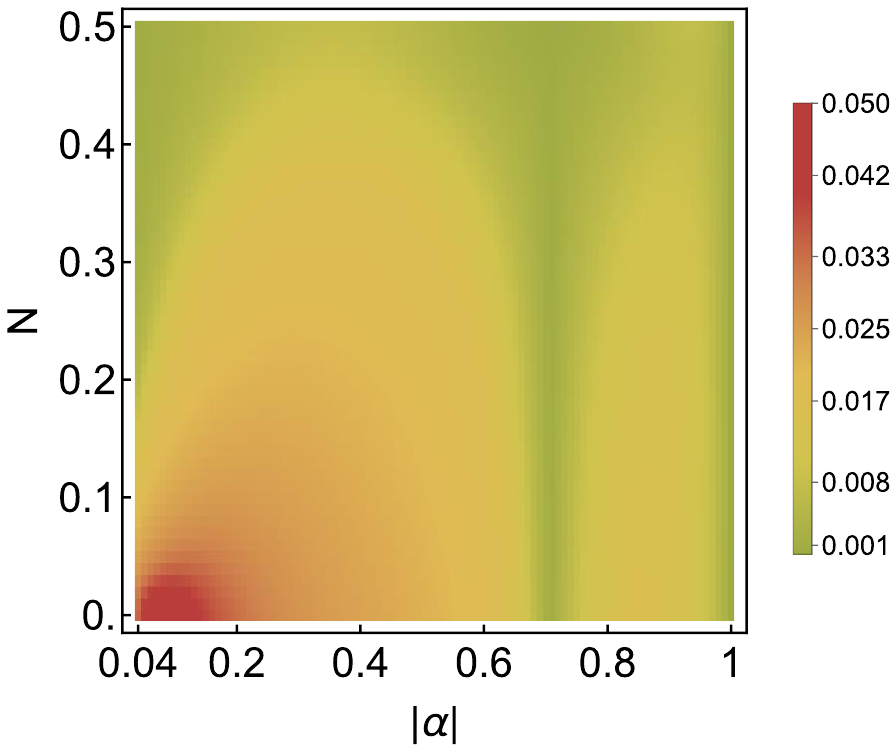}}
\caption{(a) The relationships between $R_{L}$ and $R_{P}$ with different initial states at zero temperature $N=0$. (b) The fault-tolerant(FT) phase and the non-fault-tolerant(NFT) phase. (c) The entropy variation induced by the off-diagonal elements of the density matrix under different settings. All the quantities in the figure are independent of the value of the spontaneous emission coefficient $\gamma_{0}$. This observation underscores the fact that the outcome is primarily determined by the correlation spectrum of the environment and the type of system-environment interaction.}
\label{fig: example} 
\end{figure*}

Consider the reliability loss of the logical bit, i.e. the system. Three physical bits constitute one logical bit. Each physical bit evolves independently. The master equation of the logical bits is
\begin{equation}
\begin{aligned}
        \dot{\rho}_L(t) =& \gamma_0 (1-N) \sum_{i=1}^3 \left( \sigma_{-,i} \rho_L \sigma_{+,i} - \frac{1}{2}\{\sigma_{+,i} \sigma_{-,i}, \rho_L \}
    \right) \\
    &+ \gamma_0 N \sum_{i=1}^3\left( \sigma_{+,i} \rho_L \sigma_{-,i} - \frac{1}{2} \{ \sigma_{-,i} \sigma_{+,i}, \rho_L \}
    \right),
\end{aligned}
\end{equation}
The subscript $i$ denotes the operator on the i-th qubit.
The initial state of the logical bits is $\ket{\psi_L}= \alpha \ket{111}+\sqrt{1-|\alpha|^{2}}\ket{000}$. The logical bit survives if at most one of the three physical bits is in error. The corresponding survival projector is $E_L=\ketbra{\psi_L}+\sum_{i=1}^3 X_i \ketbra{\psi_L} X_i$, where $X_i$ is the Pauli-X matrix on the $i$-th bit . The survival trajectory of the logical bit is $\ketbra{\psi_L} \to E_L \to E_L \to \cdots$. The reliability of the system results
\begin{equation}
  \begin{aligned}
         R_L(t)\overset{\delta t\rightarrow 0}{=}e^{M_1 \gamma_0 t} (\cosh{
        [M_2 \gamma_0 t]} + M_3\sinh{[M_2 \gamma_0 t]}
        )
  \end{aligned}
\end{equation}
where $M_1,\ M_2$ and $M_3$ are  time-independent constants and are determined by the initial state $\ket{\psi_L}$. A detailed derivation can be found in Supplemental materials IV\cite{Supplementarymaterial}.


The expression $\gamma_0 t = -\ln{(R_P)}/M$ allows one to establish the correlations between $R_L$ and $R_P$, as illustrated in Fig.~\ref{fig: QuantumMemoryWithFaultToleranceStructure} for different stored states at zero temperature($N=0$). When the state of the system is in $\ket{\psi_L}=\ket{111}$, it exhibits a particular structure function that can be decomposed using local Boolean algebras (blue dotted curve). The gray diagonal line $R_L=R_P$ serves as a benchmark to evaluate whether the code can enhance reliability.


If there exists a threshold $0\leq r_{c}<1$ such that $\forall R_{P}>r_{c}$, one has $R_{L}\geq R_{P}$, the system is classified as fault-tolerant (FT); otherwise, it is non-fault-tolerant (NFT). 
The presence of NFT behavior is an apparent result according to the error-correction theory, particularly in the context of the three-flip code as a classical code. This code is only capable of safeguarding systems against bit-flip errors. Here, we  provide a fresh perspective, highlighting the non-classical nature of the reliability structure function due to quantum coherence, which has a pronounced impact on stability. This leads to the classification of different settings into two phases, namely NFT and FT,  as depicted in Fig.~\ref{fig: Phase}.


As demonstrated in the preceding section, the density matrix of the apparatus exhibits off-diagonal elements. This could influence the statistical inferences of the density matrix, such as the entropy-based approach\cite{Maximumentropyapproachtoreliability}.
We introduce the difference $S_s-S_v$ between the Von-Neumann entropy $S_v\equiv-\mathrm{Tr} [\rho^{\mathrm{M}}\ln \rho^{\mathrm{M}}]$  and the Shannon entropy $S_s\equiv- \sum_{k}\rho^{\mathrm{M}}_{kk}\ln \rho^{\mathrm{M}}_{kk}$ for the classical lifetime distribution  in Fig.~\ref{fig: DeltaEntropy}. 
Comparing the two entropy, one can evaluate how much the quantum coherence could influence the inference when $\rho^{\mathrm{M}}$ involves off-diagonals.
It is evident that quantum coherence has an impact on both the non-classical structure function and the reliability-loss process, potentially exerting a substantial influence on quantum reliability. The roles of coherence in reliability-loss may vary across different quantum devices, with coherence in this example resulting in decreased reliability. As quantum coherence in much essential in quantum devices, further investigation could be an issue for quantum engineering.

\emph{Conclusions}.---
With the advancements in quantum technology, there is an increasing number of quantum devices and devices that require consideration of quantum effects. In practical applications, evaluating the reliability of such devices becomes a vital issue.

The present study establishes a universal framework for reliability theory that encompasses both classical and quantum systems. Furthermore, this study elucidates the system structure functions for quantum devices with a precise definition, offering a target function for optimizing the structures of such devices. Through examples, we observe the impact of quantum effects on system structure and reliability, which can be accurately quantified by the proposed quantum reliability. This effect is related to the redundancy allocation problem, a topic of interest in reliability engineering\cite{Mathematical_Theory_of_Reliability,devi2022review}.

In the field of quantum engineering and related technologies in quantum information and computation, this work offers a fresh perspective by shifting the focus from state-distinguishing (fidelity) to trajectory-distinguishing (reliability) measures. Specifically, quantum reliability is defined as a measure of the difference between a given trajectory and the actual quantum process in the time domain. The proposed reliability measure offers a proper extension of fidelity, enabling a more precise description of reliability loss induced by non-ideal quantum processes.

This work establishes a bridge between reliability and quantum physics, facilitating the application of expertise in quantum physics to the field of reliability engineering, thereby establishing the groundwork for analyzing the reliability of quantum devices. The proposed quantum reliability has the potential to assist in various quantum engineering technologies, including optimal design, operation, and maintenance of quantum systems.

\emph{Acknowledgements}.---
The authors appreciate Hui Dong for his advice on writing. This work was supported by the National Natural Science Foundation of China (NSFC) (Grants No. 12088101) and NSAF No. U2330401.

\bibliography{ShortBib}

\end{document}



\title{Supplemental Material for: Quantum Reliability}

\author{L.X.Cui}
\affiliation{Beijing Computational Science Research Center, Beijing, 100193, China}

\author{Y-M.Du}
\email{ymdu@gscaep.ac.cn}
\affiliation{Graduate School of China Academy of Engineering Physics, Beijing, 100193, China}

\author{C.P.Sun}
\email{suncp@gscaep.ac.cn}
\affiliation{Graduate School of China Academy of Engineering Physics, Beijing, 100193, China}
\affiliation{Beijing Computational Science Research Center, Beijing, 100193, China}
\affiliation{School of Physics, Peking University, Beijing, 100871, China}

\date{\today}

\maketitle

\section{Reliability state}
\label{App: reliability state}
In a classical system, the system's physical state is described by a point $\mathbf{x} = (x_1,x_2,\cdots ) \in \mathcal{S}$ in the \emph{physical space} $\mathcal{S}$, with $x_i$ representing the attribute of the system. In reliability analysis, we focus on the reliability states of a system, such as a binary system: survival and failure, which are represented by $1$ and $0$. A system is survival is defined as a system whose physical state is in a particular subspace $\mathcal{S}_1$ of the state space, which is called the \emph{survival subspace}. The complementary space of the reliable subspace $\mathcal{S}_0 \equiv \mathcal{S}_1^\perp$ represents the region corresponding to the system failure. The indicator function $\Sigma$ is used to describe the reliability state of the system, i.e.
\begin{equation}
  \Sigma(\mathbf{x})=
  \begin{cases}
    0,& \mathbf{x}\in \mathcal{S}_0\\
    1, & \mathbf{x} \in \mathcal{S}_1
  \end{cases}.
\end{equation}
The continuous physical state quantities are transformed into discrete reliability states. It is worth mentioning that system reliability here refers to the ability of the system to perform a specific function.

For a quantum system, its state is described by the quantum state $\ket{\psi}$. The property $E$ of the system is defined as survival, i.e., the system is said to be survival(reliable) if it possesses the property $E$, and the system fails if it does not have the property $E$. Similar to classical systems, the property $E$ of a quantum system corresponds to a subspace $\mathcal{E}$, i.e., the \emph{survival subspace}, in the entire Hilbert space $\mathcal{H}$. In quantum mechanics, whether a system state is in a subspace $\mathcal{E}$ or not can be described by the corresponding projector $\hat{E}$. A system is said to possess the property $E$ if $\ket{\psi}$ is an eigenstate of the projector $E$, i.e, $E \ket{\psi} = c \ket{\psi}$, $c$ is the corresponding eigenvalue.
Let the projector corresponding to the failure be $E^{\perp}$, and its failure subspace be $\mathcal{E}^\perp$. One can write
\begin{equation}
 E+E^{\perp} = I, \ \ \ \ \mathcal{E}+\mathcal{E}^\perp = \mathcal{H}.
\end{equation}
The states that are neither an eigenstate of $E$ nor of $E^{\perp}$ are undefined under the current context.

The above discussion is to divide the system into two states of reliability and failure, that is, the binary system. One can also make a finer division of the system state according to the practice problem, i.e., the multi-state system (MSS). In Hilbert spaces, one can find a set of orthogonal projectors to decompose the identity,
\begin{equation}
  I = \sum_i E_i,
\end{equation}
where $E_i$ corresponds to the different reliable states of the system.

\section{Trajectory weight with Markov approximation}
In the open quantum system\cite{breuer2002opensystems}, the Markov approximation is that the evolution of the system has no effect on the environment. Thus, the total density matrix of the system and environment can be written in a product form of the system state $\rho_{S}$ and environment state $\rho_{B}$ at any moment, i.e., $\rho_{S}\otimes \rho_{B}$.

Consider a system with initial state $\ket{\psi}$. 
Suppose the survival projector of the system at any moment is $E_S$ and the evolution of the system and environment is the same in every time interval for $U$. The weight of the survival trajectory $\mathcal{R}_k$ is
\begin{equation}
  W(\mathcal{R}_k)=\mathrm{Tr}[\cdots EUEU(\ketbra{\psi}\otimes \rho_B)U^\dagger E U^\dagger E\cdots],
\end{equation}
where $E=E_S\otimes I_B$ is the projector of the system and environment. According to the Markov approximation, the above equation can be simplified as
\begin{equation}\label{eq25}
    W(\mathcal{R}_k)=\mathrm{Tr}[(P_S \circ \Lambda)^N (\ketbra{\psi})\otimes \rho_B],
\end{equation}
where $P_S(\rho)=E_S\rho E_S^\dagger$ is the survival projector of the system and $\Lambda$ is the time evolution of the system.





\section{Weight of mixed trajectories }

The pure trajectories defined with the projections $E(0)\otimes E(1)\otimes\cdots E(f)$ could be extended to mixed trajectories, i.e., a series of semi-positive definite hermitian operators   $\mathcal{Y}_{\mathrm{mix}}\equiv Q(0)\otimes Q(1)\otimes\cdots Q(f)$ with $\mathrm{Tr}Q(i)$ equals to a positive integer $n_{i}$ for $i=1,2,\cdots,f$.
 This is achieved by  introducing an auxiliary system $A$ to extend the operators $Q(i)$ to some projections $\epsilon(i)$ such that $\mathrm{Tr}_{A}\epsilon(i)=Q(i)$. Since such extension is not unique, 
  for a particular trajectory, we define its weight as the largest of the various purification, reads,
 \begin{equation}\label{eq8}
W(\mathcal{Y_{\mathrm{mix}}};T)\equiv \sup_{\epsilon(0)\otimes\epsilon(1),\cdots}W(\epsilon(0)\otimes\epsilon(1)\otimes\cdots ;U\otimes I_{\mathrm{A}}),
  \end{equation}
 where the supremum is taken over all the extensions and $I_{\mathrm{A}}$ denote the identity in the space of auxiliary system.
 Specifically, if $n_{i}=1$, $\epsilon(i)$ is the purification of $Q(i)$.

\section{Calculation of the reliability for the logic qubit}
The three physical bits constitute one logical bit. Each physical bit evolves independently. The master equation of the logical bits is
\begin{equation}
\begin{aligned}
        \dot{\rho}_L(t) =& \gamma_0 (1-N) \sum_{i=1}^3 \left( \sigma_{-,i} \rho_L \sigma_{+,i} - \frac{1}{2}\{\sigma_{+,i} \sigma_{-,i}, \rho_L \}
    \right) \\
    &+ \gamma_0 N \sum_{i=1}^3\left( \sigma_{+,i} \rho_L \sigma_{-,i} - \frac{1}{2} \{ \sigma_{-,i} \sigma_{+,i}, \rho_L \}
    \right).
\end{aligned}
\end{equation}

The initial state of the logical bits is $\ket{\psi_L}= \beta \ket{0_L}+\alpha \ket{1_L}\equiv  \beta\ket{000} + \alpha \ket{111}$. The logical bit survives if at most one of the three physical bits is in error. The corresponding survival projector is $E_L=\ketbra{\psi_L}+\sum_{i=1}^3 X_i \ketbra{\psi_L} X_i$, where $X_i$ is the Pauli-X gate in the $i$-th bit . The survival trajectory of the logical bit is $\ketbra{\psi_L} \to E_L \to E_L \to \cdots$. 

Similar to the calculation of the physical qubit in the previous subsection, Let $\Lambda_L(\rho_L)\equiv \rho_L +  \dot{\rho_L} \delta t$, and $\tilde{\rho}_L(t) = a(t) \ketbra{\psi_L} + b(t) \sum_i X_i\ketbra{\psi_L}X_i$. 
The state after projection is 
\begin{equation}
    \begin{aligned}
        E_L\Lambda_L(\tilde{\rho}_L) E_L = 
        \big[
            a &- \delta t \gamma_0 (1-N) (3a\ev{\sigma_{+} \sigma_{-}} - 3b \ev{\sigma_{-}\sigma_{+}}^2)\\
            &-\delta t \gamma_0 N (3a\ev{\sigma_{-} \sigma_{+}} - 3b \ev{\sigma_{+} \sigma_{-}}^2)
        \big] \ketbra{\psi_L}\\
        +\big[
            b&+ \delta t \gamma_0 (1-N) \big(2 b|\ev{\sigma_{-}}|^2 +a\ev{\sigma_{+}\sigma_{-}}^2 -2b\ev{\sigma_{+}\sigma_{-}} - b\ev{\sigma_{-}\sigma_{+}}\big)\\
            &+\delta t \gamma_0 N \big(2b|\ev{\sigma_{+}}|^2+a\ev{\sigma_{-}\sigma_{+}}^2 -2b\ev{\sigma_{-}\sigma_{+}} - b\ev{\sigma_{+}\sigma_{-}} \big)
        \big]\sum_i X_i \ketbra{\psi_L} X_i,
    \end{aligned}
\end{equation}
where $\ev{A} = \mel{\psi_0}{A}{\psi_0}$.
This resulting to the differential equations with the initial conditions $a(0)=1,\ b(0)=0$:
\begin{equation}
    \begin{aligned}
        a'(t) =& -3\gamma_0 \left[ (1-N) (a\ev{\sigma_{+} \sigma_{-}} - b \ev{\sigma_{-}\sigma_{+}}^2)
            + N (a\ev{\sigma_{-} \sigma_{+}} - b \ev{\sigma_{+}\sigma_{-}}^2)\right]\\
        b'(t) =& \gamma_0 (1-N) (2 b|\ev{\sigma_{-}}|^2 + a\ev{\sigma_{+} \sigma_{-}}^2 -2b\ev{\sigma_{+}\sigma_{-}}-b\ev{\sigma_{-}\sigma_{+}})\\
            &+ \gamma_0 N(2 b|\ev{\sigma_{+} }|^2 +a\ev{\sigma_{-} \sigma_{+}}^2 -2b\ev{\sigma_{-}\sigma_{+}}-b\ev{\sigma_{+}\sigma_{-}}).
    \end{aligned}
\end{equation}

Here, we can get the reliability of the logical bit
\begin{equation}
  \begin{aligned}
         R_L(t)
         &={\rm Tr}[(P_L \circ \Lambda_L)^{t/\delta t} \ketbra{\psi_L}]
         \overset{\delta t\rightarrow 0}{=} {\rm Tr} [\tilde{\rho_L}] 
         = a(t)+3b(t)\\
         &=e^{M_1 \gamma_0 t} (\cosh{
        [M_2 \gamma_0 t]} + M_3\sinh{[M_2 \gamma_0 t]}
        )
  \end{aligned}
\end{equation}
where
\begin{equation}
    \begin{aligned}
        M_1&=\frac{1}{2}(q_1 + q_4),\\
        M_2&=\frac{1}{2}\sqrt{4q_2q_3 + (q_1-q_4)^2},\\
        M_3&=\frac{q_1+6q_3-q_4}{2M_2} ,\\
\end{aligned}
\end{equation}
and 
\begin{equation}
    \begin{aligned}
        q_1 =& -3(1-N) \ev{\sigma_{+}\sigma_{-}} - 3N\ev{\sigma_{-}\sigma_{+}},\\
        q_2 =& 3(1-N) \ev{\sigma_{-}\sigma_{+}}^2 + 3N \ev{\sigma_{+}\sigma_{-}}^2,\\
        q_3 =& (1-N) \ev{\sigma_{+} \sigma_{-}}^2 + N\ev{\sigma_{-}\sigma_{+}}^2,\\
        q_4 =& (1-N)(2|\ev{\sigma_{-}}|^2 - 2\ev{\sigma_{+}\sigma_{-}} - \ev{\sigma_{-}\sigma_{+}})\\
        &+ N(2|\ev{\sigma_{+}}|^2 - 2\ev{\sigma_{-}\sigma_{+}} - \ev{\sigma_{+}\sigma_{-}}).
    \end{aligned}
\end{equation}
\subsection{sufficient condition for Eq. (8)}
We are going to show the formalism Eq. (8) in the main text holds for general channels.

We represent the infinitesimal CPTP for a single physical qubit with
\begin{equation}
\Lambda_{P}(\cdot;\delta t)= I\cdot I+\delta t A_{\mu\nu}\sigma_{\mu}\cdot\sigma_{\nu},
\end{equation}
where $\sigma_{0}=I$ and $\sigma_{i},i=1,2,3$ denote the three Pauli matrix.
The condition of CPTP brings a constrain on matrix $A$, 
\begin{equation}
\begin{split}
&\sum_{\mu} A_{\mu\mu}=0,\\
&A_{0,i}+A_{i,0}-\mathrm{i}\epsilon_{ijk}A_{jk}=0,\\
&A_{\mu\nu}=A_{\nu\mu}^{*},
\end{split}
\end{equation}
with $i,j,k=1,2,3$.

The infinitesimal CPTP for the logical qubit is
\begin{equation}
\begin{split}
\Lambda_{L}(\cdot;\delta t)=I\cdot I+\sum_{i=1}^{3}\delta t A_{\mu\nu}\sigma_{\mu}^{(i)}\cdot\sigma_{\nu}^{(i)}.
\end{split}
\end{equation}
The equality as follows is required by further calculation:
\begin{equation}
\begin{split}
&\langle\psi_{L}|\sigma_{\mu}^{(i)}|\psi_{L}\rangle=\delta_{\mu,0}+\langle\sigma_{z}\rangle\delta_{\mu,3},\\
&\langle\psi_{L}|\sigma_{\mu}^{(i)}X_{j}|\psi_{L}\rangle=\delta_{\mu,1}\delta_{i,j}-\mathrm{i}\langle\sigma_{z}\rangle\delta_{\mu,2}\delta_{i,j},\\
&\langle\psi_{L}|X_{k}\sigma_{\mu}^{(i)}X_{j}|\psi_{L}\rangle=\delta_{\mu,0}\delta_{j,k}+\delta_{\mu,1}|\epsilon_{ijk}|\langle\sigma_{x}\rangle+\delta_{\mu,2}|\epsilon_{ijk}|\langle\sigma_{y}\rangle-\delta_{\mu,3}\delta_{j,k}\langle\sigma_{z}\rangle.\\
\end{split}
\end{equation}

The process of interest here is $A_{01}=A_{02}=A_{13}=A_{23}=0$, i.e., $A$ is block-diagonal.
This type of process involves both phase and spin-flip noises.

Let a state be
\begin{equation}
\begin{split}
a|\psi_{L}\rangle\langle\psi_{L}|
+b\sum_{i}X_{i}|\psi_{L}\rangle\langle\psi_{L}|X_{i}.
\end{split}
\end{equation}
Represent $\sum_{i}  E(A_{\mu\nu}\sigma_{\mu}^{(i)}\cdot\sigma_{\nu}^{(i)}+A_{\nu\mu}\sigma_{\nu}^{(i)}\cdot\sigma_{\mu}^{(i)})E/(1+\delta_{\mu\nu})$ as a matrix $D_{\mu\nu}$ acting on this space.
\begin{equation}
\begin{split}
&D_{00}=\begin{pmatrix}3A_{00}&0\\
0&3A_{00}
\end{pmatrix}\\
&D_{03}=\begin{pmatrix}3\langle\sigma_{z}\rangle (A_{0,3}+A_{3,0})&0\\
0&\langle\sigma_{z}\rangle (A_{0,3}+A_{3,0})\\
\end{pmatrix}\\
&D_{11}=\begin{pmatrix}0&A_{11}\\
3A_{11}&2\langle\sigma_{x}\rangle^{2} A_{11}\\
\end{pmatrix}\\
&D_{12}=\begin{pmatrix}0&\mathrm{i}\langle \sigma_{z}\rangle (A_{21}-A_{12})\\
3\mathrm{i}\langle \sigma_{z}\rangle (A_{12}-A_{21})&2\langle\sigma_{y}\rangle\langle\sigma_{x}\rangle(A_{12}+A_{21})\\
\end{pmatrix}\\
&D_{22}=\begin{pmatrix}0&\langle\sigma_{z}\rangle^{2}A_{22}\\
3\langle\sigma_{z}\rangle^{2}A_{22}&2\langle\sigma_{y}\rangle^{2}A_{22}\\
\end{pmatrix}\\
&D_{33}=\begin{pmatrix}\langle \sigma_{z}\rangle^{2}A_{33}&0\\
0&3\langle \sigma_{z}\rangle^{2}A_{33}\\
\end{pmatrix}.\\
\end{split}
\end{equation}

The reliability is
\begin{equation}
R(t)=\begin{pmatrix}1&3
\end{pmatrix}\exp[\sum_{\mu,\nu}D_{\mu,\nu}^{\mathrm{T}}t]\begin{pmatrix}1\\0
\end{pmatrix}
=e^{M_1  t} (\cosh{
        [M_2  t]} + M_3\sinh{[M_2 t]})
\end{equation}
where the coefficient $M_{1},M_{2},M_{3}$ depends on $|\psi_{L}\rangle$ and $A_{\mu\nu}$.
This is the same as Eq. (10) in the main text, where $\gamma_{0}$ has been absorbed into these coefficients.

\section{Structure of apparatus density matrix}
In terms of the property of the apparatus density matrix as follows, the matrix representation for $\Lambda_{L}(\cdot)E$ will be applied to the direct calculation of the off-diagonal elements of the apparatus density matrix.

Define the notations as $\Lambda_{L,E}(\cdot)\equiv E\Lambda_{L}(\cdot )E$,
$\tilde{\Lambda}_{L}(\cdot)\equiv\Lambda_{L}(\cdot)E$
and
\begin{equation}
G_{k,n}\equiv \mathrm{Tr}\Bigl\{\tilde{\Lambda}_{L}^{\circ n}[\Lambda_{L,E}^{\circ k}(|\psi_{L}\rangle\langle\psi_{L}|)]\Bigr\}.
\end{equation}

Then, the following relations hold:
\begin{equation}
\begin{split}
G_{k,0}-G_{k+1,0}&=\rho^{\mathrm{M}}_{k,k}\\
G_{k,n}-G_{k,n+1}&=\sum_{k'\geq k}\rho^{\mathrm{M}}_{k',k+n}
\end{split}
\end{equation}
It follows that 
\begin{equation}
\begin{split}
\rho^{\mathrm{M}}_{k,k+n}=G_{k,n}-G_{k,n+1}-G_{k+1,n-1}+G_{k+1,n}
\end{split}
\end{equation}
and in the continuous limit
\begin{equation}\label{eq23}
\begin{split}
\rho^{\mathrm{M}}_{t,t+\tau}=\begin{pmatrix}1&3&0&0\end{pmatrix}\mathcal{D}_{\mathcal{L}E}\exp(\mathcal{D}_{\mathcal{L}E}\tau)\mathcal{D}_{E\mathcal{L}E}\exp(\mathcal{D}_{E\mathcal{L}E}t)\begin{pmatrix}1\\0\\0\\0\end{pmatrix}
\end{split}
\end{equation}
where $\mathcal{D}_{\mathcal{L}E}$  and $\mathcal{D}_{E\mathcal{L}E}$ denote the matrix representations of super operators $\mathcal{L}(\cdot)E$ and $E\mathcal{L}(\cdot)E$, respectively.
These two matrices are 
\begin{equation}\label{eq24}
\begin{aligned}
\mathcal{D}_{E\mathcal{L}E}=\begin{pmatrix}
    D_{1}&\mathbf{0}\\
    \mathbf{0}&\mathbf{0}
 \end{pmatrix}\\
 \mathcal{D}_{\mathcal{L}E}=\begin{pmatrix}
    D_{1}&\exp(\mathrm{i}\phi)D_{2}\\
    \exp(-\mathrm{i}\phi)D_{3}&D_{4}
 \end{pmatrix}
\end{aligned}
\end{equation}
where, $D_{1},D_{2},D_{3},D_{4}$ are  only depend on the probability $|\alpha|^{2}$, and  $\phi$ is the  phase $\exp(\mathrm{i}\phi)=\alpha^{*}\beta^{*}/|\alpha\beta|$.  
\begin{equation}
        D_1=\begin{pmatrix}
            -3(1-N)|\alpha|^2-3N|\beta|^2 & 3N |\alpha|^4 + 3(1-N) |\beta|^4\\
            (N-1)|\alpha|^4+N|\beta|^4 & -(1+N)|\beta|^2-|\alpha|^2(-2+N+2|\beta|^2)
        \end{pmatrix}
\end{equation}
\begin{equation}
    D_2= |\alpha \beta| \begin{pmatrix}
        1.5-3N &  3|\beta|^2 -3N\\
        N-|\alpha|^2  & 0.5+N -2|\beta|^2
    \end{pmatrix}
\end{equation}
\begin{equation}
    D_3= |\alpha \beta| \begin{pmatrix}
        1.5-3N &  3|\beta|^2 -3N\\
        N-|\alpha|^2  & 2.5-3N -2|\beta|^2
    \end{pmatrix}
\end{equation}
\begin{equation}
    D_4=\begin{pmatrix}
        -1.5 & 3|\alpha\beta|^2\\
        |\alpha\beta|^2 & -1.5-2|\alpha\beta|^2
    \end{pmatrix},
\end{equation}
where the coefficients satisfy $|\alpha|^{2}+|\beta|^{2}=1$.
This implies that $\rho^{\mathrm{M}}_{t,t+\tau}$ does not depend on the phase $\phi$, since the phase $\phi$ also contains the global phase that could not influence the observable quantities. And this is also verified directly as follows.
Firstly, for any matrices $D_{i},D_{i}'$ that do not depend on the phase, one has
\begin{equation}
\begin{split}
\begin{pmatrix}
D_{1}&\exp(\mathrm{i}\phi)D_{2}\\\exp(-\mathrm{i}\phi)D_{3}&D_{4}
 \end{pmatrix}
\begin{pmatrix}
D'_{1}&\exp(\mathrm{i}\phi)D'_{2}\\\exp(-\mathrm{i}\phi)D'_{3}&D'_{4} \end{pmatrix}=\begin{pmatrix}
D''_{1}&\exp(\mathrm{i}\phi)D''_{2}\\\exp(-\mathrm{i}\phi)D''_{3}&D''_{4}
 \end{pmatrix}
\end{split}
\end{equation}
where $D''_{i}$ are also phase independent.
Then we have $\forall n\in\mathbb{Z}$
\begin{equation}
\begin{split}
\begin{pmatrix}
I&\mathbf{0}\\\mathbf{0}&\mathbf{0} 
 \end{pmatrix}
 \mathcal{D}_{\mathcal{L}E}^{n}\begin{pmatrix}
I&\mathbf{0}\\\mathbf{0}&\mathbf{0} 
 \end{pmatrix}
\end{split}
\end{equation}
does not depend on $\phi$.

Thus the off-diagonal elements of the apparatus density matrix depend on the diagonal terms.

\section{Example: Shor's nine-bit code}
In its setting, the reliable projector for a logical bit is
\begin{equation}
E=\sum_{i=1}^{9}\Bigl[X_{i}|\psi_{L}\rangle\langle\psi_{L}|X_{i}+Y_{i}|\psi_{L}\rangle\langle\psi_{L}|Y_{i}\Bigr]+\sum_{i=1,4,7}Z_{i}|\psi_{L}\rangle\langle\psi_{L}|Z_{i},
\end{equation}
where the encode state $|\psi_{L}\rangle\equiv \alpha |1_{L}\rangle+\beta|0_{L}\rangle$,
with
\begin{equation}
\begin{split}
|1_{L}\rangle=(|000\rangle+|111\rangle)(|000\rangle+|111\rangle)(|000\rangle+|111\rangle)/2\sqrt{2}\\
|0_{L}\rangle=(|000\rangle-|111\rangle)(|000\rangle-|111\rangle)(|000\rangle-|111\rangle)/2\sqrt{2}
\end{split}
\end{equation}

The reliable projector for a physical bit is $|\psi_{P}\rangle\langle\psi_{P}|$ with
\begin{equation}
|\psi_{P}\rangle\equiv  \alpha |1\rangle+\beta|0\rangle
\end{equation}
According to the computational method described in the letter, the results are illustrated in Fig.\ref{fig: RofAFQ}.

\begin{figure}[htbp]
    \includegraphics[width=0.5\linewidth]{[1,3].pdf}
    \caption{The reliability for the quantum storage protected by nine-bit code. The  state to be storage are $(|1\rangle+3|0\rangle)/\sqrt{10}$ (blue solid line), $(|1\rangle+1|0\rangle)/\sqrt{2}$(red dash line) and $|1\rangle$ (black dot-dash line). The gray double-dot line denote $R_{L}=R_{P}$.}
    \label{fig: RofAFQ} 
\end{figure}

\section{Example: an adaptive feedback quantum sensing system}
We consider the scheme presented in Ref.\cite{Wiseman2000,Wiseman2001,Sanders2010}. A $N_{q}$-qubit state is initially stored in a quantum memory. The qubits are read out with photons one by one, which are injected into Mach-Zehnder interferometer and measured as the output.
The phase difference $\theta$ between the two arms of a Mach-Zehnder interferometer is estimated by introducing an additional phase shifter $\Theta_{i}$, $i=1,2,3,\cdots$. By injecting photons one by one and detecting which arm the photon ejects as the output of this step. Then, adjust the additional phase shifter properly according to the output values, the phase difference is evaluated with $\theta=\Theta_{N_{q}}$.

Now we computed the quantum reliability of the device. For a common protocol if the photon in any one step is lost, then the task is failed. Since the device accomplishes the task step by step, that is, the projector for survival should be defined at different time points. Thus quantum reliability as a trajectory-dependent measure is much more suitable to the situation.

\begin{figure*}[htbp]
    \label{fig: RofAFQ2} 
    \includegraphics[width=0.5\linewidth]{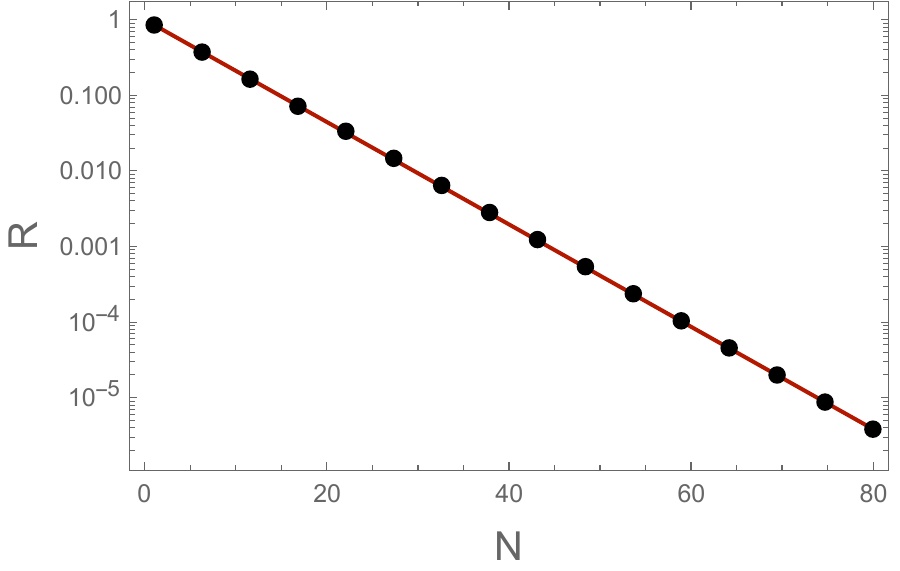}
      \includegraphics[width=0.5\linewidth]{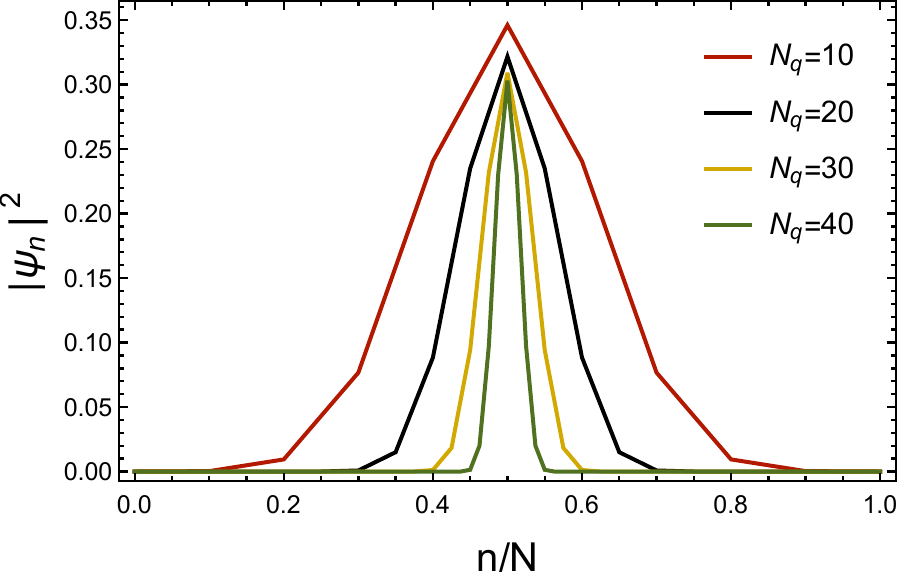}
\caption{(a)Reliability of adaptive feedback quantum sensing system versus the number of qubits with $l_{0}=0.2,l_{1}=0.1$
 ,$N_{q}=1,2,3,\cdots,80$; (b) the peak of $|\psi_{m}|^{2}$ with $N_{q}=10,20,40,80$}\label{closeloop}
\end{figure*}

If neglect the interaction between the channels, then
for the $i$-th step,  the process for the system before measurement is a unitary operation $\mathcal{U}^{(i)}$ with loss, namely $\Lambda_{(i)}=I^{(1)}\otimes I^{(2)}\otimes\cdots\otimes I^{(i-1)}\otimes\mathcal{U}^{(i)}\circ \Lambda^{(i)}_{Loss}\otimes I^{N-1}\cdots\otimes I^{(N)}$.
We depict the process of loss by introducing an additional state $|L\rangle$. Together with the one-photon state for the two modes $|0\rangle,|1\rangle$, there are three levels in every step.
Suppose the probability of a photon loss on two modes are $l_{0},l_{1}$, respectively. The process is 
\begin{equation}
\begin{split}
\Lambda^{(i)}_{Loss}(\rho^{(i)})=&(1-l_{0})\rho_{00}^{(i)}|0\rangle\langle 0|+(1-l_{1})\rho_{11}^{(i)}|1\rangle\langle 1|+(l_{0}\rho_{00}^{(i)}+l_{1}\rho_{11}^{(i)}+\rho_{LL}^{(i)})|L\rangle\langle L|\\
&+\sqrt{(1-l_{0})(1-l_{1})}\rho_{01}^{(i)}|0\rangle\langle 1|+\sqrt{(1-l_{0})}\rho_{0L}^{(i)}|0\rangle\langle L|+\sqrt{(1-l_{1})}\rho_{1L}^{(i)}|1\rangle\langle L|+\mathrm{h.c}.
\end{split}
\end{equation}
The reliable space at the $i$-th step is $E_{i}\equiv I^{(1)}\otimes I^{(2)}\otimes\cdots\otimes I^{(i-1)}\otimes [|0\rangle\langle 0|+|1\rangle\langle 1|]^{(i)}\otimes\cdots\otimes [|0\rangle\langle 0|+|1\rangle\langle 1|]^{(N)}$.
The reliability is given as
\begin{equation}
R(N)=\mathrm{Tr}\Bigl\{P_{E_{N}}\circ\Lambda_{N}\circ\cdots\circ P_{E_{1}}\circ\Lambda_{1}(|\psi\rangle\langle\psi|)\Bigr\},
\end{equation}
where $P_{E_{i}}(\cdot)\equiv E_{i}\cdot E_{i}$.
For any state $|\psi\rangle=\sum_{\mathbf{\sigma}} a_{\mathbf{\sigma}}|\mathbf{\sigma}\rangle$, 
the reliability is $\sum_{\mathbf{\sigma}} |a_{\mathbf{\sigma}}|^{2} \prod_{i} (1-l_{\sigma_{i}})$.
Previous literature\cite{Wiseman2000,Wiseman2001,Sanders2010} usually considers the state with the symmetric bases   $|\psi\rangle=\sum \psi(n)|n,N_{q}-n\rangle$,  the reliability varies with the particle number $\ln R(N_{q})\approx N_{q}\ln\lambda $. 
The coefficient $\lambda$ is evaluated as
$\lambda = (1-l_{0})^{n_{*}/N_{q}}(1-l_{1})^{1-n_{*}/N_{q}}$, where $n_{*}=\mathrm{Argmax}|\psi_{n}|^{2}$. 
For the example, we give the result of the sine state\cite{Sanders2010} in Fig. \ref{closeloop},
$\lambda=\sqrt{(1-l_{0})(1-l_{1})}$.

\bibliography{ShortBib}